\documentclass[preprintnumbers,amsmath,amssymbm,prl]{revtex4}
\usepackage{epsfig}
\usepackage{graphicx}

\begin{document}
\title{Relaxation dynamics of charged gravitational collapse}
\author{Shahar Hod}
\affiliation{The Ruppin Academic Center, Emeq Hefer 40250, Israel}
\affiliation{ } \affiliation{The Hadassah Institute, Jerusalem
91010, Israel}
\date{\today}

\begin{abstract}
\ \ \ We study analytically the relaxation dynamics of charged test
fields left outside a newly born charged black hole. In particular,
we obtain a simple analytic expression for the fundamental
quasinormal resonances of near-extremal Reissner-Nordstr\"om black
holes. The formula is expressed in terms of the black-hole physical
parameters: $\omega=q\Phi-i2\pi T_{BH}(n+{1 \over 2})$, where
$T_{BH}$ and $\Phi$ are the temperature and electric potential of
the black hole, and $q$ is the charge of the field.
\end{abstract}
\bigskip
\maketitle


The radiative test fields of complete gravitational collapse decay
with time leaving behind a `bald' black hole. This is the essence of
the no-hair conjecture introduced by Wheeler in the early 1970's
\cite{Wheeler}. The various no-hair theorems state that the external
field of a dynamically formed black hole (or a perturbed black hole)
relaxes to a Kerr-Newman spacetime, characterized solely by the
black-hole mass, charge, and angular momentum. This implies that
test fields left outside the newly born black hole would either be
radiated away to infinity or swallowed by the black hole.

The relaxation phase in the dynamics of perturbed black holes is
characterized by `quasinormal ringing', damped oscillations with a
discrete spectrum (see e.g. \cite{Nollert1} for a review and
detailed lists of references). At late times, all perturbations
(test fields) are radiated away in a manner reminiscent of the last
pure dying tones of a ringing bell \cite{Press,Cruz,Vish,Davis}. The
black hole quasinormal modes (QNMs) correspond to solutions of the
perturbations equations with the physical boundary conditions of
purely outgoing waves at spatial infinity and purely ingoing waves
crossing the event horizon. Such boundary conditions single out a
{\it discrete} set of black-hole resonances $\{\omega_n\}$ (assuming
a time dependence of the form $e^{-i\omega t}$).

In accord with the spirit of the no-hair conjecture \cite{Wheeler},
the external test fields would either fall into the black hole or
radiate to infinity. This implies that the perturbation (test field)
decays with time and the corresponding QNM frequencies are therefore
{\it complex}. It turns out that there exist an infinite number of
quasinormal modes, characterizing oscillations with decreasing
relaxation times (increasing imaginary part), see
\cite{Leaver,Hodln3,KeshHod} and references therein. The mode with
the smallest imaginary part, known as the fundamental mode,
determines the characteristic dynamical timescale for generic
perturbations to decay \cite{Ono,Hod1,Hod2,Gruz,Hodb,Hod8,Hod9}. [We
recall that, in asymptotically flat spacetimes the quasinormal
ringing is followed by an inverse power-law decaying tail. However,
it should be noted that most of the energy of the test field is
radiated away to infinity (or swallowed by the black hole) in the
quasinormal ringing phase \cite{Nollert1}. The decaying tail
contains only a small fraction of the field's energy. It is
therefore physically justified to determine the characteristic
lifetime of the perturbation (of the test field) from the decay of
its associated QNMs.]

In this work we study the nearly spherical gravitational collapse of
charged matter to form a charged Reissner-Nordstr\"om (RN) black
hole. The black-hole formation is followed by a relaxation phase
which describes the decay of two types of perturbation fields (test
fields):\ (1) coupled gravitational-electromagnetic test fields,
and\ (2) charged test fields. In the present work we focus on the
relaxation dynamics of the {\it charged} test fields left outside
the newly born black hole. In particular, we shall determine the
fundamental (least-damped) resonant frequencies of a {\it charged}
scalar field in the charged black-hole spacetime. As we shall show
below, the spectrum of charged quasinormal resonances can be studied
analytically in the near-extremal limit $Q\to M$ (we use natural
units for which $G=c=\hbar=1$), where $M$ and $Q$ are the mass and
charge of the black hole, respectively. [This also corresponds to
the $T_{BH}\to 0$ limit, where
$T_{BH}={{(M^2-Q^2)^{1/2}}\over{2\pi[M+(M^2-Q^2)^{1/2}]^2}}$ is the
Bekenstein-Hawking temperature of the black hole.]

The dynamics of a charged, massive scalar field $\Psi$ in the RN
spacetime is governed by the Klein-Gordon equation
\begin{equation}\label{Eq1}
[(\nabla^\nu-iqA^\nu)(\nabla_{\nu}-iqA_{\nu}) -\mu^2]\Psi=0\  ,
\end{equation}
where $A_{\nu}$ is the electromagnetic potential. Here $q$ and $\mu$
are the charge and mass of the field, respectively. [$q$ and $\mu$
stand for $q/\hbar$ and $\mu/\hbar$ respectively, and they have the
dimensions of $($length$)^{-1}$.] One may decompose the field as
\begin{equation}\label{Eq2}
\Psi_{lm}(t,r,\theta,\phi)=e^{im\phi}S_{lm}(\theta)R_{lm}(r)e^{-i\omega
t}\ ,
\end{equation}
where $(t,r,\theta,\phi)$ are the standard Schwarzschild
coordinates, $\omega$ is the (conserved) frequency of the mode, $l$
is the spherical harmonic index, and $m$ is the azimuthal harmonic
index with $-l\leq m\leq l$. (We shall henceforth omit the indices
$l$ and $m$ for brevity.) With the decomposition (\ref{Eq2}), $R$
and $S$ obey radial and angular equations both of confluent Heun
type coupled by a separation constant $A$ \cite{Heun,Flam}.

The angular functions $S(\theta)$ are the familiar spherical
harmonics which are solutions of the angular equation \cite{Flam}
\begin{equation}\label{Eq3}
{1\over {\sin\theta}}{\partial \over
{\partial\theta}}\Big(\sin\theta {{\partial
S}\over{\partial\theta}}\Big)+\Big(A-{{m^2}\over{\sin^2\theta}}\Big)S=0\
.
\end{equation}
The angular functions are required to be regular at the poles
$\theta=0$ and $\theta=\pi$. These boundary conditions yield the
well-known set of eigenvalues $A_l=l(l+1)$, where $l$ is an integer.

The radial Klein-Gordon equation is given by \cite{HodPirpam}
\begin{equation}\label{Eq4}
\Delta{{d} \over{dr}}\Big(\Delta{{dR}\over{dr}}\Big)+\Big[K^2
-\Delta[\mu^2r^2+l(l+1)]\Big]R=0\ ,
\end{equation}
where $\Delta\equiv r^2-2Mr+Q^2$ and $K\equiv \omega r^2-qQr$. The
zeroes of $\Delta$, $r_{\pm}=M\pm (M^2-Q^2)^{1/2}$, are the black
hole (event and inner) horizons.

In order to determine the free resonances of the field one should
impose the physical boundary conditions of purely ingoing waves at
the black-hole horizon (as measured by a comoving observer) and
purely outgoing waves at spatial infinity. That is,
\begin{equation}\label{Eq5}
R \sim
\begin{cases}
{1\over r}e^{i\sqrt{\omega^2-\mu^2}y} & \text{ as }
r\rightarrow\infty\ \ (y\rightarrow \infty)\ ; \\
e^{-i (\omega-q\Phi)y} & \text{ as } r\rightarrow r_+\ \
(y\rightarrow -\infty)\ ,
\end{cases}
\end{equation}
where the ``tortoise" radial coordinate $y$ is defined by
$dy=(r^2/\Delta)dr$ and $\Phi=Q/r_+$ is the electric potential at
the horizon. These boundary conditions single out a discrete set of
resonances $\{\omega_n\}$ which correspond to free oscillations of
the field. These resonances determine the ringdown response of a
black hole to external perturbations. Note that the boundary
conditions of outgoing waves at spatial infinity imply $\omega>\mu$.

It is convenient to define new dimensionless variables
\begin{equation}\label{Eq6}
x\equiv {{r-r_+}\over {r_+}}\ \ ;\ \ \tau\equiv{{r_+-r_-}\over
{r_+}}\ \ ;\ \ \varpi\equiv{{\omega-q\Phi}\over{2\pi T_{BH}}}\ \ ;\
\ k\equiv 2\omega r_+-qQ\  ,
\end{equation}
in terms of which the radial equation becomes
\begin{equation}\label{Eq7}
x(x+\tau){{d^2R}\over{dx^2}}+(2x+\tau){{dR}\over{dx}}+VR=0\  ,
\end{equation}
where $V\equiv K^2/r^2_+x(x+\tau)-[\mu^2r^2_+(x+1)^2+l(l+1)]$ and
$K=r^2_+\omega x^2+r_+kx+r_+\varpi\tau/2$. As we shall now show, the
radial equation is amenable to an analytic treatment in the regime
$\tau\ll 1$ with $M(\omega-q\Phi)\ll 1$.

We first consider the radial equation (\ref{Eq7}) in the far region
$x\gg \text{max}\{\tau,M(\omega-q\Phi)\}$. Then Eq. (\ref{Eq7}) is
well approximated by
\begin{equation}\label{Eq8}
x^2{{d^2R}\over{dx^2}}+2x{{dR}\over{dx}}+V_{\text{far}}R=0\  ,
\end{equation}
where $V_{\text{{far}}}=(\omega^2-\mu^2)r^2_+x^2+2(\omega
k-\mu^2r_+)r_+x-[\mu^2r^2_++l(l+1)-k^2]$. A solution of Eq.
(\ref{Eq8}) that satisfies the boundary condition (\ref{Eq5}) can be
expressed in terms of the confluent hypergeometric functions
$M(a,b,z)$ \cite{Morse,Abram}
\begin{equation}\label{Eq9}
R=C_1(2i\sqrt{\omega^2-\mu^2}r_+)^{{1\over 2}+i\delta}x^{-{1\over
2}+i\delta}e^{-i\sqrt{\omega^2-\mu^2}r_+x}M({1\over
2}+i\delta+i\kappa,1+2i\delta,2i\sqrt{\omega^2-\mu^2}r_+x)+C_2(\delta\to
-\delta)\  ,
\end{equation}
where $C_1$ and $C_2$ are constants and
\begin{equation}\label{Eq10}
\delta^2\equiv k^2-\mu^2r^2_+-(l+{1\over 2})^2\ \ ;\ \ \kappa\equiv
{{\omega k-\mu^2r_+}\over{\sqrt{\omega^2-\mu^2}}}\ .
\end{equation}
The notation $(\delta\to -\delta)$ means ``replace $\delta$ by
$-\delta$ in the preceding term."

We next consider the near horizon region $x\ll 1$. The radial
equation is given by Eq. (\ref{Eq7}) with $V\to
V_{\text{near}}\equiv-[\mu^2r^2_++l(l+1)]+(kx+\varpi\tau/2)^2/x(x+\tau)$.
The physical solution obeying the ingoing boundary conditions at the
horizon is given by \cite{Morse,Abram}
\begin{equation}\label{Eq11}
R=x^{-{i\over 2}\varpi}\Big({x\over \tau}+1\Big)^{i({1\over
2}\varpi-k)}{_2F_1}({1\over 2}+i\delta-ik,{1\over
2}-i\delta-ik;1-i\varpi;-x/\tau)\  ,
\end{equation}
where $_2F_1(a,b;c;z)$ is the hypergeometric function.

The solutions (\ref{Eq9}) and (\ref{Eq11}) can be matched in the
overlap region $\text{max}\{\tau,M(\omega-q\Phi)\}\ll x\ll 1$. The
$x\ll 1$ limit of Eq. (\ref{Eq9}) yields \cite{Morse,Abram}
\begin{equation}\label{Eq12}
R\to C_1(2i\sqrt{\omega^2-\mu^2}r_+)^{{1\over 2}+i\delta}x^{-{1\over
2}+i\delta}+C_2(\delta\to -\delta)\  .
\end{equation}
The $x\gg \tau$ limit of Eq. (\ref{Eq11}) yields \cite{Morse,Abram}
\begin{equation}\label{Eq13}
R\to \tau^{{1\over
2}-i\delta-i\varpi/2}{{\Gamma(2i\delta)\Gamma(1-i\varpi)}\over{\Gamma({1\over
2}+i\delta-ik)\Gamma({1\over 2}+i\delta-i\varpi+ik)}}x^{-{1\over
2}+i\delta}+(\delta\to -\delta)\  .
\end{equation}
By matching the two solutions in the overlap region one finds
\begin{equation}\label{Eq14}
C_1=\tau^{{1\over
2}-i\delta-i\varpi/2}{{\Gamma(2i\delta)\Gamma(1-i\varpi)}\over{\Gamma({1\over
2}+i\delta-ik)\Gamma({1\over
2}+i\delta-i\varpi+ik)}}(2i\sqrt{\omega^2-\mu^2}r_+)^{-{1\over
2}-i\delta}\  ,
\end{equation}
\begin{equation}\label{Eq15}
C_2=\tau^{{1\over
2}+i\delta-i\varpi/2}{{\Gamma(-2i\delta)\Gamma(1-i\varpi)}\over{\Gamma({1\over
2}-i\delta-ik)\Gamma({1\over
2}-i\delta-i\varpi+ik)}}(2i\sqrt{\omega^2-\mu^2}r_+)^{-{1\over
2}+i\delta}\  .
\end{equation}

Approximating Eq. (\ref{Eq9}) for $x\to\infty$ one gets
\cite{Morse,Abram}
\begin{eqnarray}\label{Eq16}
R&\to&
\Big[C_1(2i\sqrt{\omega^2-\mu^2}r_+)^{i\kappa}{{\Gamma(1+2i\delta)}\over{\Gamma({1\over
2}+i\delta+i\kappa)}}x^{-1+i\kappa}+C_2(\delta\to
-\delta)\Big]e^{i\sqrt{\omega^2-\mu^2}r_+x}\nonumber
\\&& + \Big[C_1(2i\sqrt{\omega^2-\mu^2}r_+)^{-i\kappa}{{\Gamma(1+2i\delta)}\over{\Gamma({1\over
2}+i\delta-i\kappa)}}x^{-1-i\kappa}(-1)^{-{1\over
2}-i\delta-i\kappa}+C_2(\delta\to
-\delta)\Big]e^{-i\sqrt{\omega^2-\mu^2}r_+x}\ .
\end{eqnarray}
A free oscillation (a quasinormal resonance) is characterized by a
purely outgoing wave at spatial infinity. The coefficient of the
ingoing wave $e^{-i\sqrt{\omega^2-\mu^2}r_+x}$ in Eq. (\ref{Eq16})
should therefore vanish. Taking cognizance of Eqs.
(\ref{Eq14})-(\ref{Eq16}), one finds the resonance condition for the
free oscillations of the field:
\begin{equation}\label{Eq17}
{{\Gamma(2i\delta)\Gamma(1+2i\delta)(-2i\tau\sqrt{\omega^2-\mu^2}r_+)^{-i\delta}}\over{\Gamma({1\over
2}+i\delta-i\kappa)\Gamma({1\over 2}+i\delta-ik)\Gamma({1\over
2}+i\delta-i\varpi+ik)}}+{{\Gamma(-2i\delta)\Gamma(1-2i\delta)(-2i\tau\sqrt{\omega^2-\mu^2}r_+)^{i\delta}}\over{\Gamma({1\over
2}-i\delta-i\kappa)\Gamma({1\over 2}-i\delta-ik)\Gamma({1\over
2}-i\delta-i\varpi+ik)}}=0\  .
\end{equation}
The resonance condition (\ref{Eq17}) can be solved analytically in
the regime $\tau\ll 1$ with $\omega\simeq q\Phi$; We first write it
in the form
\begin{equation}\label{Eq18}
{{1}\over{\Gamma({1\over 2}-i\delta-i\varpi+ik)}}={\cal D}\times
(-2i\tau\sqrt{\omega^2-\mu^2}r_+)^{-2i\delta}\ ,
\end{equation}
where ${\cal D}\equiv [\Gamma(2i\delta)]^2\Gamma({1\over
2}-i\delta-i\kappa)\Gamma({1\over
2}-i\delta-ik)/[\Gamma(-2i\delta)]^2\Gamma({1\over
2}+i\delta-i\kappa)\Gamma({1\over 2}+i\delta-ik)\Gamma({1\over
2}+i\delta-i\varpi+ik)$. We note that ${\cal D}$ has a well defined
limit as $Q\to M$ and $\omega\to q\Phi$.

In the limit $\omega\to q\Phi$, where $\omega$ is almost purely
real, one finds from Eq. (\ref{Eq10}) that $\delta^2$ is also almost
purely real. As we shall now show, the resonance condition Eq.
(\ref{Eq18}) can be solved analytically if $\delta\gtrsim 1$. Thus,
we restrict our attention to the regime $q^2>\mu^2+O(M^{-2})$. Then
one has
$(-i)^{-2i\delta}=e^{(-i{{\pi}\over{2}})(-2i\delta)}=e^{-\pi\delta}\ll
1$. One therefore finds $\epsilon\equiv
(-2i\tau\sqrt{\omega^2-\mu^2}r_+)^{-2i\delta}\ll 1$.

Thus, a consistent solution of the resonance condition (\ref{Eq18})
may be obtained if $1/\Gamma({1\over
2}-i\delta-i\varpi+ik)=O(\epsilon)$ \cite{Notedelta}. Suppose
\begin{equation}\label{Eq20}
{1\over 2}-i\delta-i\varpi+ik=-n+\eta\epsilon+O(\epsilon^2)\ ,
\end{equation}
where $n\geq 0$ is a non-negative integer and $\eta$ is a constant
to be determined below. Then one has
\begin{equation}\label{Eq21}
\Gamma({1\over
2}-i\delta-i\varpi+ik)\simeq\Gamma(-n+\eta\epsilon)\simeq
(-n)^{-1}\Gamma(-n+1+\eta\epsilon)\simeq\cdots\simeq [(-1)^n
n!]^{-1}\Gamma(\eta\epsilon)\  ,
\end{equation}
where we have used the relation $\Gamma(z+1)=z\Gamma(z)$
\cite{Abram}. Next, using the series expansion
$1/\Gamma(z)=\sum_{k=1}^{\infty} c_k z^k$ with $c_1=1$ [see Eq.
$(6.1.34)$ of \cite{Abram}], one obtains
\begin{equation}\label{Eq22}
1/\Gamma({1\over 2}-i\delta-i\varpi+ik)=(-1)^n
n!\eta\epsilon+O(\epsilon^2)\ .
\end{equation}
Substituting (\ref{Eq22}) into (\ref{Eq18}) one finds $\eta={\cal
D}/[(-1)^n n!]$.

Finally, substituting $\varpi=(\omega-q\Phi)/2\pi T_{BH}$ and
$k=qQ+O(MT_{BH})$ for $\omega=q\Phi+O(T_{BH})$ into Eq.
(\ref{Eq20}), one obtains a simple formula for the quasinormal
resonances of the charged field:
\begin{equation}\label{Eq23}
\omega={{qQ}\over{r_+}}-i2\pi T_{BH}[(n+{1\over
2}-\eta\epsilon)-i(\delta-qQ)]\ ,
\end{equation}
where $\delta=[(q^2-\mu^2)Q^2-(l+1/2)^2]^{1/2}+O(MT_{BH})$ and
$n=0,1,2,...$\ .

As emphasized above, our boundary conditions require
$\Re\omega>\mu$. From Eq. (\ref{Eq23}) one finds
$\Re\omega=q[1+O(MT_{BH})]$ in the near-extremal limit $MT_{BH}\to
0$. Thus, our analytical treatment is valid in the regime
$q/\mu>1+O(MT_{BH})$.

Taking cognizance of Eq. (\ref{Eq23}) one finds
$\Im\omega=O(T_{BH})$ for the {\it charged} test fields in the
near-extremal limit. We note that numerical studies have indicated
that {\it neutral} test fields (coupled
gravitational-electromagnetic perturbations and Dirac perturbations)
are characterized by $\Im\omega=O(M^{-1})$ \cite{Jingn}. One
therefore realizes that the relaxation dynamics of the newly born
charged black hole is dominated by the charged perturbation (test)
fields (Note that $T_{BH}\ll M^{-1}$ in the near-extremal limit.) We
would like to emphasize that our analytical approximation is valid
for {\it real} values of $\delta$ with $\delta\gtrsim 1$. This
implies that we {\it cannot} extrapolate our results to the case of
neutral test fields: Had we taken in Eq. (\ref{Eq10})
$\Re\omega\simeq q\Phi\to 0$ for the neutral test fields, we would
have found that $\delta$ is {\it imaginary} in this limit-- this
would be beyond the regime of validity of our analytical
approximation.

In summary, we have analyzed the relaxation dynamics of charged
gravitational collapse. In particular, we have studied analytically
the quasinormal mode spectrum of charged fields in the spacetime of
a near-extremal charged black hole. It was shown that the
fundamental resonances can be expressed in terms of the black-hole
physical parameters: the temperature $T_{BH}$ and the electric
potential $\Phi$.

It is worth mentioning that a fundamental bound on the relaxation
time $\tau_{\text{relax}}$ of a perturbed thermodynamical system has
recently been suggested \cite{Hod1,Hod2,Gruz,Pesci,Hodb,Hod8,Hod9},
$\tau_{\text{relax}} \geq \hbar/\pi T$, where $T$ is the system's
temperature. This bound can be regarded as a quantitative
formulation of the third law of thermodynamics. Taking cognizance of
this relaxation bound, one may deduce an upper bound on the
black-hole fundamental (slowest damped) frequency
\begin{equation}\label{Eq24}
\min\{\Im\omega\} \leq \pi T_{BH}/\hbar\  .
\end{equation}
Thus the relaxation bound implies that a black hole should have (at
least) one quasinormal resonance whose imaginary part conform to the
upper bound (\ref{Eq24}). This mode would dominate the relaxation
dynamics of the newly born black hole and will determine its
characteristic relaxation timescale. Taking cognizance of Eq.
(\ref{Eq23}) and substituting $n=0$ for the fundamental quasinormal
resonance, one obtains $\min\{\Im\omega\}=\pi T_{BH}/\hbar$. One
therefore concludes that near-extremal charged black holes actually
saturate the universal relaxation bound.

\bigskip
\noindent
{\bf ACKNOWLEDGMENTS}
\bigskip

This research is supported by the Meltzer Science Foundation. I
thank Liran Shimshi, Clovis Hopman, Yael Oren, and Arbel M. Ongo for
helpful discussions.



\begin{thebibliography}{99}
\bibitem{Notedelta} Had we taken $\delta<0$, we would have found
that $\epsilon\gg 1$. In this case a consistent solution of the
resonance condition (\ref{Eq18}) would require
$1/\Gamma(1/2+i\delta-i\varpi+ik)=O(\epsilon^{-1})$. It is
straightforward to show that the black-hole resonances would still
be given by the same analytical formula, see Eq. (\ref{Eq23}) below.
\end{thebibliography}
\end{document}